%% file: euro.tex
\documentstyle[graphics]{europhys}
\input euromacr

\begin{document}
\euro{}{}{}{}
\Date{}
\shorttitle{L.M. JENSEN {\it et al.}: DYNAMIC CRITICAL BEHAVIORS OF 3D
$XY$ MODELS ETC.}
\title{Dynamic critical behaviors of three-dimensional
$XY$ models related to superconductors/superfluids}
\author{Lars Melwyn Jensen, Beom Jun Kim, \And Petter Minnhagen}
\institute{Department of Theoretical Physics, Ume{\aa} University,
901 87 Ume{\aa}, Sweden} 
\rec{}{}
\pacs{
\Pacs{74}{40$+$k}{Fluctuations (noise, chaos, nonequilibrium superconductivity, localization, etc.)}
\Pacs{05}{70Jk}{Critical point phenomena}
\Pacs{75}{40Gb}{Dynamic properties (dynamic susceptibility, spin waves, spin
         diffusion, dynamic scaling, etc.) }
}
\maketitle
\begin{abstract}
The dynamic critical exponent $z$ is determined from
numerical simulations
for the
three-dimensional $XY$ model subject to two types of dynamics,
{\it i.e.},
relaxational dynamics and resistively shunted junction (RSJ) dynamics,
as well as for two different treatments of the boundary,
 {\it i.e.}, periodic boundary
condition (PBC) and fluctuating twist boundary condition (FTBC).
In case of relaxational dynamics, finite size scaling at the critical
temperature gives $z\approx 2$ for PBC and $1.5$ for
FTBC, while for RSJ dynamics $z\approx 1.5$ is obtained in both
cases. The results are discussed in the context of
superfluid/superconductors and
vortex dynamics, and are compared with what have been found
for other related models.
\end{abstract}

A neutral superfluid like $^4$He and
a superconductor 
in the limit of large London penetration depth can be characterized by a
complex order parameter, and
the $XY$ model can be viewed as a discretized version
of this type of systems~\cite{minnhagen_rev}.
All these systems are expected to belong to the same
universality class for the thermodynamic critical properties of the phase
transition. An interesting feature of these models is the
presence of thermally generated topological defects.
In two dimensions (2D) the topological defects take the form of
vortices and give rise
to the
Kosterlitz-Thouless transition~\cite{minnhagen_rev,kosterlitz}.
Also in 3D
thermally generated vortex loops are present at
the transition and it has been argued that the critical
properties, both the static and the dynamic,
can be associated with these vortex loops~\cite{williams1,williams2,williams3}.
The low-temperature phase in the 3D case consists of closed vortex loops of
finite extent whereas in the high-temperature phase the loops can disintegrate~\cite{williams1,williams2,williams3}.

In the present Letter we investigate the dynamic critical properties
for two simple cases when the static properties are given by the 3D $XY$ model.
One connection between the vortex loops and the dynamical properties
is through the $2\pi$ phase slip across the system occurring when a
vortex loop expands so much that it leaves the system. The connection
is most easily phrased in case of a superconductor: the rate at which
the vortex loops expands and leaves the system, when driven by a
dc current, is proportional to the voltage across the
sample. Consequently the vortex loops are connected to the resistance
which for a system in equilibrium can be obtained by the
fluctuation-dissipation theorem\cite{weber}.

Similarly to the static case one expects universality also
for the dynamic critical properties in the sense that the critical
dynamics does not depend on the details but rather on general characteristics 
like conservation laws, spatial dimensions, and the static critical
properties~\cite{hohenberg}.
The motion of the vortex loops are associated with a conservation law
since the vorticity of a fixed area can only
change by vortex segments leaving or entering the area.  
This conservation law restricts the motion of the vortex loops and
consequently one may expect that the longest relaxation time of the
system can be associated with the vortex loops.
In the dynamic universality
classes defined by Hohenberg and Halperin~\cite{hohenberg} the dynamics of a 3D
superfluid belongs
to model F characterized by the dynamic critical exponent $z=3/2$\cite{notez}.
A model with purely relaxational dynamics on the other hand belongs to
model A with $z\approx 2$~\cite{hohenberg}. 
    
The two cases we study are relaxational dynamics
which does not have local current conservation, and the resistively shunted junction
(RSJ) dynamics which has local current conservation.
The Hamiltonian for the 3D $XY$ model on an $L\times L\times L $ cubic
lattice can be expressed as
\begin{equation} \label{eq_H}
H = -J \sum_{\langle i j\rangle} \cos(\theta_i - \theta_j - {\bf r}_{ij} \cdot {\bf \Delta}),
\end{equation}
where the sum is over all nearest-neighbor pairs,
$\theta_i - \theta_j - {\bf r}_{ij} \cdot {\bf \Delta}$ is the
difference in the spin direction between the neighboring sites
$i$ and $j$, and $J$ is the coupling
strength. The twist variable ${\bf
\Delta}=(\Delta_x,\Delta_y,\Delta_z)$ is a vector such that $L\Delta_x$
measures the average rotation of the spin direction when going from
one boundary
surface to the opposite in the $x$ direction and similarly for the other
directions,
${\bf r}_{ij}$ is the unit vector from site $i$ to the
nearest-neighbor
site $j$ (the lattice spacing is taken to be unity), and 
$\theta_i$ is a phase angle associated with each site $i$ measured
with respect
to the local spin-direction associated with a uniform twist
${\bf \Delta}$ across the sample.
We use the boundary conditions 
$\theta_i=\theta_{i+L{\bf \hat x}}=\theta_{i+L{\bf \hat y}}=\theta_{i+L{\bf \hat z}}$.

In the 
superfluid/superconductor analogy of the $XY$ model, $\theta_i -
\theta_j - {\bf r}_{ij} \cdot {\bf \Delta}$ is 
the total (gauge invariant)
phase difference of the order parameter and the
twist
$\bf {\bf \Delta}$ may be thought of as the contribution to
the gauge-invariant phase from a spatially uniform vector
potential.
 
The usual periodic
boundary conditions (PBC) for the $XY$ model
correspond to periodic spin directions and   
to ${\bf \Delta}=0$. However this imposes an unphysical restriction
on the topological defects (the vortex loops): the
original defect-free state is not regained with PBC, when a defect is
created in a defect-free state and then annihilated across the
boundary~\cite{olsson}.
The additional degrees of freedom introduced by
${\bf \Delta}$, on the other hand, ensure that the energy associated with a given configuration
of topological defects is unique~\cite{olsson}.
The more physical boundary condition which includes the fluctuations
of the twist
is termed the fluctuating twist boundary condition (FTBC)~\cite{olsson,beom}.

In the RSJ case the total current $i_{ij}$ from  $i$ to $j$ is the sum
of the supercurrent, the normal resistive current, and a thermal noise current:
\begin{equation}\label{eq_ilink}
i_{ij} = i_c \sin(\theta_i - \theta_j -{\bf r}_{ij}\cdot {\bf \Delta}) + \frac{V_{ij}}{r} + \eta_{ij},
\end{equation}
where $i_c \equiv 2e J/ \hbar$ is the critical current of a single junction, 
$V_{ij}$
is the potential difference across the junction,  $r$ is the 
shunt resistance.
The current conservation law at each site, together with the Josephson
 relation 
$d(\theta_i - \theta_j-{\bf r}_{ij}\cdot {\bf \Delta} )/dt = 2eV_{ij}/\hbar$, allows us to write the
 equations of motion in the form
\begin{equation}\label{eq_rsj}
\dot\theta_i = -\sum_j G_{ij}{\sum_k}^{'} [\sin(\theta_j -
\theta_k-{\bf r}_{jk}\cdot {\bf \Delta}) + \eta_{jk} ],
\end{equation}
where the primed summation is over six nearest neighbors of $j$, $G_{ij}$ is the lattice 
Green function on the cubic lattice. For convenience we from
now on use units such that $i_c=J=r=\hbar/2e=1$.
The remaining dynamical equation for ${\bf \Delta}$ is
 obtained from the local current conservation together with the
global current conservation condition that no currents pass through
the boundaries~\cite{beom}:
\begin{equation} \label{eq_dot_delta}
\dot{\bf \Delta } = -\Gamma_{\Delta} 
\frac{\partial H}{\partial {\bf \Delta}} + \eta_{\bf \Delta}
\end{equation}
with $\Gamma_\Delta=1/L^3$.
In order to ensure the correct thermal equilibrium the noise
correlations obey the relations: 
$\langle \eta_{ij}(t) \rangle = 0$,
$\langle \eta_{ij}(t) \eta_{kl} (0)\rangle =
2T(\delta_{ik}\delta_{jl} -  \delta_{il}\delta_{jk})\delta(t)$,
and correspondingly for the three components of
$\eta_{\bf \Delta}$:
$\langle\eta_{\Delta_m}(t) \rangle = 0$,
$\langle \eta_{\Delta_m}(t)\eta_{\Delta_n}(0)  \rangle = 
(2T/L^3)\delta_{mn} \delta (t)$ with $m,n=x,y,z$.

The RSJ equations defined in this way incorporates local current conservation and
the boundary conditions are chosen such that
the vorticity for each of the six sides of the cubic lattice is zero
at any instant and that there is no current flow through them. The RSJ
equations
are usually phrased in the superconducting language,
however, they also apply to a neutral superfluid; the RSJ equations in this
case correspond to a constant mass density and local conservation of mass
current. 

We also
note that eq.~(\ref{eq_dot_delta}), which is related
to the resistance and hence to the vortex loops, has a 
relaxational form. However, one should note that the relaxational
constant $\Gamma_\Delta= L^{-3}$ is unusual since it vanishes with the
size of the system and that eq.~(\ref{eq_dot_delta}) by itself can be
viewed
as a global current conservation law expressing that 
the average total current
of the system vanishes at each instant.

How important is strict local current conservation for 
the critical dynamics?
We investigate this by comparing the results from the RSJ dynamics to
relaxational dynamics, where eq.~(\ref{eq_rsj})
is replaced by
the purely relaxational form:
\begin{equation}\label{eq_tdgl}
\frac{d \theta_i (t)}{dt} = -\Gamma \frac{\partial H}{\partial \theta_i} + \eta_i (t)    
\end{equation}
with $\langle \eta_i (t) \rangle = 0$ and
$\langle \eta_i (t) \eta_j (0) \rangle = 2T \delta_{ij}\delta(t)$ (we
have set $\Gamma\equiv 1$). Thus the dynamics is in this case given by the
two relaxational equations (\ref{eq_dot_delta}) and (\ref{eq_tdgl}).
Superficially one might have guessed that this relaxational 
dynamics
should belong to model A\cite{hohenberg}. However, as will be shown below,
the $z$ value
obtained from size scaling at $T_c$ is not compatible with this 
expectation. This suggests that the global conservation law reflected
in the size dependent relaxation constant in eq.~(\ref{eq_dot_delta})
is enough to slow down the critical dynamics. 

The dynamical equations are integrated numerically using the second
order algorithm in ref.~\cite{batrouni} with a discrete time
step $\Delta t=0.05$ for RSJ and $\Delta t=0.05$ and $0.01$ for
relaxational dynamics, using lattice sizes up to $L=32$ and $24$, respectively.

The resistance $R$ is related to the equilibrium fluctuations of ${\bf \Delta}(t)$ by the
fluctuation-dissipation theorem~\cite{beom}:  
\[
R = \frac{L^2}{2 T } \frac{1}{\Theta} \langle [ \Delta_m(\Theta) - \Delta_m(0) ]^2 \rangle ,
\]
where $\Theta$ is a large enough time interval  
($\Theta= 2000$ in the present simulation). 
Near the second order phase transition the intensive quantity $LR(T,L)$
obeys the scaling relation:
\begin{equation}\label{eq_rscale}
  LR(T,L)=L^{-(z-1)}\tilde{\rho}[L^{1/\nu}(T-T_c)] .
  \end{equation}
Thus if we plot the ratio $\ln [R(T,L)/R(T,L')]/\ln(L/L')$ as a function
of $T$ for different pairs $(L,L')$ then the curves should cross at
($T_c$, $-z$)~\cite{lidmar}. The inset in fig.~1 shows
the RSJ result for the pairs $(L,L')=$$(4,8)$, $(8,16)$,
and $(4,16)$. The crossing point gives $T_c=2.20$, which is very close
to the true critical temperature $T_c \approx 2.202$ for the 3D
$XY$ model~\cite{gottlob}, and $z=1.46$.
Figure~1 confirms this determination by aid of the full scaling relation
eq.~(\ref{eq_rscale})
using $T_c$ and $z$ obtained above together with
$\nu=0.67$
for the 3D $XY$ model ($\nu\approx 0.671\pm 0.001$~\cite{olsson1}). As seen a very
good scaling
collapse is obtained. We have estimated the precision in the
determination by
treating $z$ and $T_c$ as free parameters in the full scaling
relation with the result $z=1.46\pm 0.06$.
One reason for treating $T_c$ as a free variable is that in principle
the finite time step $\Delta t$
in the integration introduces an uncertainty in $T_c$~\cite{beom}.
Figure 2 gives the corresponding result for relaxational dynamics.
In this case the integration turns out to be more sensitive to the
choice of $\Delta t$.
In order to handle this we
calculate $R$ at $T_c=2.20$ where $R\propto L^{-z}$ [see
eq.~(\ref{eq_rscale})]
with $\Delta t=0.05$ and $0.01$, 
extrapolate
linearly to $\Delta t=0$, and obtain $z\approx 1.5$.

\begin{figure}
\begin{center}
\resizebox*{!}{6.5cm}{\includegraphics{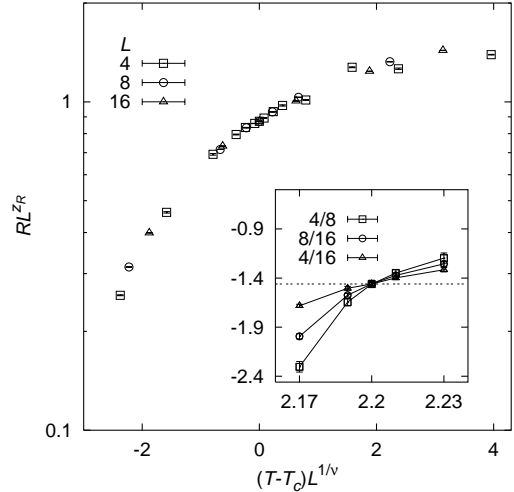}}
\end{center}
\caption{The scaling curve for resistance $R$ in the RSJ case. The
    parameters used are $z=1.46$ and $T_c=2.20$, determined by
    the intersection construction described in the text and shown in
    the inset, together with the value $\nu=0.67$ expected for the 3D
    $XY$ model. From these scaling constructions we obtain the estimate
    $z=1.46\pm 0.06$.}
\label{fig_rsj_R}
\end{figure}

\begin{figure}
\begin{center}
\resizebox*{!}{6.5cm}{\includegraphics{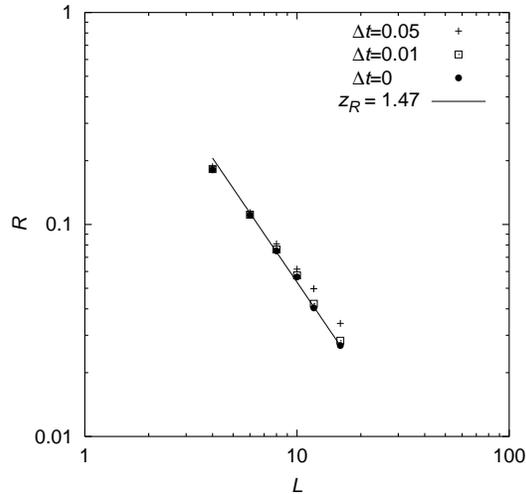}}
\end{center}
\caption{Determination of the dynamical critical exponent $z$ for the
  relaxational dynamics with FTBC from resistance scaling $R\propto
    L^{-z}$ at $T_c=2.20$. The data obtained for $L=4$, $8$, $10$,
    $12$, and $16$ and the two integration time
    steps $\Delta t=0.05$ and $0.01$ were linearly extrapolated to
    $\Delta t=0$. The value $z=1.47$ was obtained from a least square fit.}
\label{fig_rsj_psi}
\end{figure}

Our conclusion is that RSJ dynamics and relaxational dynamics have the
same size scaling of the resistance at the critical temperature and
that the $z$ value obtained from this scaling is
consistent with $z \approx 1.5$. This implies that the constraint imposed by the
local current
conservation is less crucial than one might have thought.
We suggest the following hand-waving explanation: Equation~(\ref{eq_tdgl})
describes individual spins relaxing   
towards a state with a given value of ${\bf \Delta}$.
Each fixed configuration of vortex loops correspond to a
twist ${\bf \Delta}$. Since ${\bf \Delta}$ has a slow relaxation governed by
eq.~(\ref{eq_dot_delta}) this suggests that the dynamics is
compatible with a situation
where the change of a vortex loop configuration is slow compared to the 
relaxation of the individual spins. From this perspective RSJ dynamics
and the relaxational
dynamics described above are just two alternative ways of imposing a slow dynamics on the vortex loops.

Next we consider the case when ${\bf \Delta}=0$ which 
corresponds to the standard PBC imposed on the spins. Relaxational dynamics
is in this case given by eq.~(\ref{eq_tdgl}) with ${\bf \Delta}=0$.
This is compatible with the situation when the spins are relaxing
directly towards
the global ground state with ${\bf \Delta}=0$.
In this case we cannot use eq.~(\ref{eq_rscale}) to find $z$ (because ${\bf \Delta}=0$).
Instead we resort to the following size scaling relation valid at $T_c$\cite{melwyn,note}:
\begin{equation} \label{gscale}
G(t)=\frac{1}{L}h(tL^{-z}) , 
\end{equation}
where $G(t)=\langle F(t)F(0)\rangle/L^3$ is the time-correlation
function, with 
$F(t)=\sum_{\langle ij \rangle_x}\sin(\theta_i - \theta_j)$ 
and the sum is over all links in one direction\cite{beom}.
In the superconductor analogy this is related to the supercurrent
correlations\cite{beom}.
Figure 3 shows that a good scaling is obtained for $z=2$.
From this we conclude that relaxational dynamics with standard PBC
has $z\approx 2$ consistent with the model A universality class\cite{hohenberg,dorsey}.
Our suggested interpretation is that, when the average twist ${\bf \Delta}$ is not
one of the dynamical variables, the global current constraint reflected in 
eq.~(\ref{eq_dot_delta}) is no longer present and
the relaxational dynamics becomes of standard relaxational type.
We compare this to the RSJ dynamics for the same case, {\it i.e.},
standard PBC for which ${\bf \Delta}=0$. Figure 4 shows that a
good scaling collapse
is obtained for $z=1.5$, which is the same as for FTBC. 
This suggests that local current conservation is a sufficient but not
a necessary condition for imposing the slow vortex loop
dynamics. 

\begin{figure}
\begin{center}
\resizebox*{!}{6.5cm}{\includegraphics{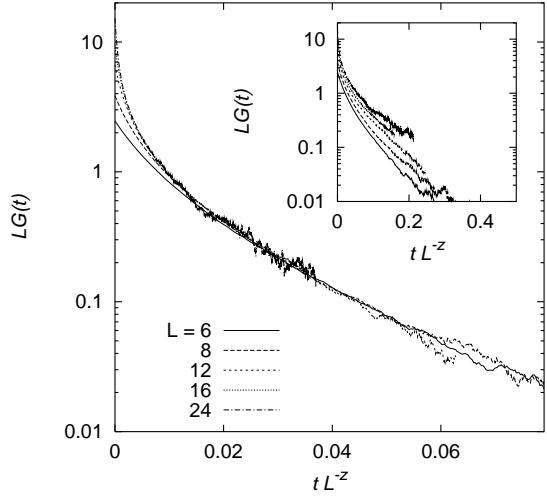}}
\end{center}
\caption{Determination of the $z$ for relaxational dynamics with PBC
  from the scaling function eq.~(\ref{gscale}) $LG(t)=h(tL^{-z})$: A
  good scaling collapse is obtained for $z=2$. The inset shows that no
  scaling collapse is obtained for $z=1.5$.}
\label{fig_tdgl_R}
\end{figure}

\begin{figure}
\begin{center}
\resizebox*{!}{6.5cm}{\includegraphics{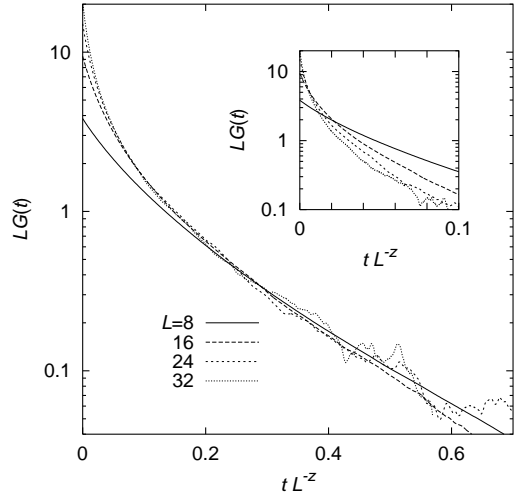}}
\end{center}
\caption{Determination of the dynamical critical exponent $z$ for
    the RSJ case with PBC from the scaling function eq.~(\ref{gscale}) $LG(t)=h(tL^{-z})$:
    A good scaling collapse is obtained for $z=1.5$. The inset shows
    that no scaling collapse is obtained for $z=2$.}
\end{figure}

The suggestion, that the dynamics of the vortex loops
can be associated
with an exponent $z\approx 1.5$, can be further substantiated in the
following way:
The Hamiltonian for the $XY$ model with FTBC is dual to 
the lattice vortex loop model with PBC\cite{olsson}.
This means that relaxational dynamics for the $XY$ model with FTBC
corresponds to relaxational dynamics for the lattice vortex loop model with PBC.
In the latter model the only degrees of freedom are the vortex loops
and
it has been shown that $z\approx 1.5$ is obtained from size scaling
of the resistance for this
model\cite{weber,lidmar}.
From this perspective it is tempting to conclude that $z\approx 1.5$
can be
associated with pure relaxational dynamics for the vortex loops.  
In addition one may note that an exponent $z\approx 1.5$ is also
consistent with $z=1.44$ in ref.~\cite{williams1} obtained
from a theoretical treatment of vortex loops. 

Finally we note that in ref.~\cite{stroud} the value $z=1.5\pm 0.5$ was
determined for the RSJ model in the presence of external currents and that
in ref.~\cite{landau} 
$z=1.38\pm 0.05$ was found from simulations of a version of the $XY$ model
with spin dynamics which is an alternative dynamics consistent with superfluids.
For the lattice vortex loop model with Monte Carlo dynamics 
$z=1.45\pm 0.05$ was obtained in ref.~\cite{lidmar}
and $z=1.51\pm 0.03$ in ref.~\cite{weber}
using the same method as in the inset of fig.~1.

This leaves us with the following two main alternatives: $z$ determined
from size scaling for 3D $XY$ model
with relaxational dynamics and FTBC, the lattice vortex loop model with
relaxational dynamics and PBC, as well as the 3D $XY$ model with RSJ
dynamics in all cases gives the value $z=3/2$ corresponding to model F. This
would then be different from the 3D $XY$ model with spin
dynamics in ref.~\cite{landau} and the vortex loop prediction in
ref.~\cite{williams1}. The other possibility is that all cases
correspond to a vortex loop dynamics with a $z$ slightly lower than
3/2. Our present precision is not enough to distinguish between these alternatives.

In conclusion we have from simulations determined the dynamic critical exponent $z$
by using size scaling at $T_c$ for
the 3D $XY$ model with relaxational and RSJ dynamics. For relaxational dynamics
with PBC we obtain $z\approx 2$ which is consistent with model A 
dynamics~\cite{hohenberg,dorsey}. However 
we conclude  that the relaxational dynamics with FTBC, the lattice vortex loop model  with
relaxational dynamics and PBC\cite{weber}, as well as RSJ dynamics with both PBC
and FTBC
all have the value $z\approx 1.5$. We suggest that the reason for this
agreement
is that all these models effectively corresponds  to relaxational
dynamics
of the vortex loops. Model F corresponds to $z=3/2$ which is consistent
with our result although the slightly lower value $z=1.44$ for vortex
loops in ref.~\cite{williams1} is also consistent.

\stars
This work was supported by the Swedish Natural Research Council
through contract FU 04040-332.

\end{document}

%% file: euromacr.tex
\def\And{{\rm and\ }}

\def\stars{\bigskip\centerline{***}\medskip}

\newif\ifboo \boofalse

\def\Review#1{\boofalse{\it #1},}
\def\Name#1{{\sc #1},}
\def\Vol#1{\ifboo Vol. {\bf #1}\else{\bf #1}\fi}
\def\Year#1{\ifboo #1\else(#1)\fi}

\def\Page#1{\ifboo {\rm p. #1}\else{\rm #1}\fi}

%% file: euro.bbl
\begin{thebibliography}{99}
\bibitem{notez} The equality holds for a model where the critical
  exponent of the specific heat has a value $\alpha\leq 0$
  as for the 3D $XY$ model.
\bibitem{note} This $G(t)$ scaling cannot be used for FTBC because
the helicity modulus at zero wavevector is always zero\cite{melwyn}.
\end{thebibliography}
